\documentclass[trackchanges, twocolumn, twocolappendix,resetfootnote]{aastex701}

\usepackage{amsmath}
\begin{document}

\title{Optimizing Prompt Radio Counterpart Searches for Gravitational-Wave Events with the Deep Synoptic Array}

\renewcommand{\sectionautorefname}{Section}

\newcommand{\affilone}{Cahill Center for Astronomy and Astrophysics, California Institute of Technology, Pasadena, CA 91125, USA}
\newcommand{\affiltwo}{Owens Valley Radio Observatory, California Institute of Technology, Big Pine, CA 93513, USA}
\newcommand{\affilthree}{LIGO Laboratory, California Institute of Technology, Pasadena, CA 91125, USA}

\newcommand{\affilassign}[1]{%
    \ifnum#1=1 \affiliation{\affilone}\fi
    \ifnum#1=2 \affiliation{\affiltwo}\fi
    \ifnum#1=3 \affiliation{\affilthree}\fi
    \ifnum#1=4 \affiliation{\affilfour}\fi
    \ifnum#1=5 \affiliation{\affilfive}\fi
    \ifnum#1=6 \affiliation{\affilsix}\fi
    \ifnum#1=7 \affiliation{\affilseven}\fi
    \ifnum#1=8 \affiliation{\affileight}\fi
    \ifnum#1=9 \affiliation{\affilnine}\fi
    \ifnum#1=11 \affiliation{\affilelev}\fi
    \ifnum#1=12 \affiliation{\affiltwelve}\fi
    \ifnum#1=13 \affiliation{\affilthirteen}\fi
}

\author[0000-0003-1226-118X]{Nikita Kosogorov}
\email[show]{nakosogorov@gmail.com} 
\email[show]{nkosogor@caltech.edu} 
\affilassign{1}
\affilassign{2}

\author[0009-0001-4899-9955]{Max Melching}
\email{mmelchin@caltech.edu} 
\affilassign{3}

\author[0000-0002-7083-4049]{Gregg Hallinan}
\email{gh@astro.caltech.edu}
\affilassign{1}
\affilassign{2}

\author[0000-0001-8115-0577]{Jonah Kanner}
\email{jkanner@caltech.edu}
\affilassign{3}

\author[0000-0002-0928-6784]{Alan J. Weinstein}
\email{ajw@caltech.edu}
\affilassign{3}

\author[]{Kevin Lacker}
\email{lacker@gmail.com}
\affilassign{1}
\affilassign{2}

\author[0000-0002-7252-5485]{Vikram Ravi}
\email{vikram@astro.caltech.edu}
\affilassign{1}
\affilassign{2}

\begin{abstract} 
Prompt millisecond-duration coherent radio emission is proposed as a counterpart to neutron-star-containing compact mergers, produced before, during, or shortly after coalescence. Its detection would probe the plasma environment, magnetic fields, remnant evolution, and emission physics of these mergers. At GHz frequencies, dispersive delays can range from fractions of a second to several seconds, requiring the telescope to be on source before or soon after coalescence. Negative- or low-latency alerts, rapid response, and optimized pointing under uncertain, evolving localizations are therefore essential. We present a framework for prompt gravitational-wave follow-up with the upcoming Deep Synoptic Array (DSA), a sensitive, fast-slewing, wide-field GHz radio array. We simulate low-latency localization sequences and embed them in a follow-up environment with elevation, field-of-view, and slew constraints. We compare a reinforcement-learning pointing policy with a baseline that slews toward the highest-probability accessible region at each step. This peak-tracking rule is a strong, interpretable baseline for initial deployment, while the learned policy provides a complementary strategy for a subset of broad, multimodal, or arc-shaped localizations. At a random merger time, $37\%$ of events are observable above $20^\circ$ elevation; among these, $\sim90\%$ of BNS events are successfully reached and tracked using zero-noise localization maps, decreasing to $\sim75\%$ with Gaussian-noise maps. We validate the optimized pointing trajectories on prototype DSA antennas. DSA follow-up can reach a fluence sensitivity of $\sim\!20\,{\rm mJy\,ms}$ over a representative 700--1025\,MHz sub-band, probing millisecond bursts with luminosities of $\sim\!10^{38}$--$10^{39}\,{\rm erg\,s^{-1}}$, at or below the lower end of several coherent-emission predictions for neutron-star mergers.

\end{abstract}



\section{Introduction}

The direct detection of gravitational waves (GW) has made compact-object mergers accessible as multimessenger astrophysical sources.  For binary neutron star mergers, GW170817 provided the defining example \citep{2017PhRvL.119p1101A,2017ApJ...848L..12A}, linking the GW signal to a rich electromagnetic counterpart: optical and infrared observations revealed a kilonova powered by freshly synthesized $r$-process material \citep[e.g.,][]{2017Sci...358.1556C,2017ApJ...848L..17C,2017Natur.551...67P,2017Natur.551...80K,2017Sci...358.1559K,2017Natur.551...75S,2017Sci...358.1565E}; high-energy and radio observations showed that the merger produced a relativistic jet associated with a short gamma-ray burst \citep[e.g.,][]{2017ApJ...848L..14G,2017ApJ...848L..15S,2017Sci...358.1579H,2017Natur.551...71T,2018ApJ...856L..18M,2018ApJ...863L..18A,2018Natur.561..355M}; and modeling of the broadband emission constrained the viewing geometry, improving the standard-siren measurement of the Hubble constant \citep[e.g.,][]{2017Natur.551...85A,2018Natur.561..355M,2018ApJ...863L..18A,2019NatAs...3..940H,2024PhRvD.109f3508P}.  These results relied on follow-up spanning the electromagnetic spectrum, including radio emission observed on afterglow timescales well after the merger.

One predicted counterpart, however, remains elusive: prompt coherent radio emission associated with neutron-star containing mergers. Such emission may occur before coalescence, during the merger itself, or in the seconds to minutes that follow. Binary neutron star (BNS) and neutron star--black hole (NSBH) systems are particularly interesting in this context because they can involve a magnetized neutron star and therefore allow several channels for coherent radio production.

A wide range of mechanisms has been proposed for such emission. In the pre-merger stage, interaction between the magnetospheres of the binary components may generate coherent radiation through magnetic reconnection, unipolar-inductor-like processes, or related plasma effects \citep[e.g.,][]{1996A&A...312..937L,2001MNRAS.322..695H,2012ApJ...755...80P,2016ApJ...822L...7W,2013PhRvL.111f1105P,2020PhRvD.101f3017C,2021PhRvD.104f3004C,2015ApJ...814L..20M,2020ApJ...893L...6M,2023PhRvL.130x5201M}. Resonant tidal excitation and crust shattering provide an additional precursor channel, in which crustal oscillations launch Alfvén waves that evolve nonlinearly in the magnetosphere and may power radio emission \citep{2012PhRvL.108a1102T,2024ApJ...973L..37M}. During or immediately after merger, a short-lived hypermassive or supramassive neutron-star remnant, or the collapse of such a remnant, may power additional coherent emission channels \citep[e.g.,][]{2011MNRAS.413.2031M,2012PhRvD..86j4035L,2013PASJ...65L..12T,2014A&A...562A.137F,2014MNRAS.441.2433R}. On somewhat longer but still prompt timescales, shocks associated with relativistic outflows or strongly magnetized winds may also produce coherent radio bursts \citep{2000A&A...364..655U}, including GHz fast radio burst (FRB)-like signals from synchrotron-maser emission in magnetar-driven or merger-driven shocks \citep{2014MNRAS.442L...9L,2021MNRAS.501.3184S}. Recent analytic and numerical studies further support the plausibility of coherent merger-time radio emission, including scenarios in which even moderately strong magnetic fields can generate bright transients near coalescence \citep[e.g.,][]{2020ApJ...890L..24Z,2020ApJ...893L...6M,2023ApJ...956L..33M,2023MNRAS.519.3923C,2025PhRvD.111h3023C}.

Despite the substantial uncertainty in the underlying emission physics, the scientific payoff of a detection would be considerable. A prompt radio counterpart would directly connect at least part of the coherent-radio transient population to compact mergers, helping to clarify whether some FRB-like events can arise in this channel \citep[e.g.,][]{2013PASJ...65L..12T, 2014A&A...562A.137F,2018PASJ...70...39Y,2020ApJ...890L..24Z}. A tentative association between an FRB and a BNS merger has also been discussed, although such associations remain uncertain \citep{2023NatAs...7..579M}. Its propagation properties could also probe the source environment: the dispersion measure may constrain the local plasma column, while scattering, Faraday rotation, and polarization could provide information on the circumburst medium and magnetic field structure near the source \citep[e.g.,][]{2019ARA&A..57..417C, 2018ApJ...861..150P,2018Natur.553..182M,2023PhRvD.108d3009K}. Finally, the arrival time of any radio burst relative to the GW signal would constrain the emission mechanism itself, helping to distinguish between precursor, merger-time, delayed-collapse, and post-merger scenarios. In delayed-collapse models, such a measurement could also constrain the remnant lifetime and provide complementary information on the neutron-star equation of state \citep[e.g.,][]{2014MNRAS.441.2433R,2014PhRvD..89d7302L}. For these reasons, prompt radio emission from BNS and NSBH mergers remains a well-motivated target for rapid follow-up searches.

This opportunity is closely tied to alert latency. For neutron-star mergers, the most interesting signals may be millisecond-duration bursts emitted within seconds of coalescence, including before the final merger in precursor scenarios. Plasma dispersion delays their arrival at radio frequencies, potentially providing a brief window for GW-triggered follow-up. For dispersion measures of $100$--$1000~{\rm pc\,cm^{-3}}$, the propagation delay relative to infinite frequency is approximately $0.4$--$4$~s at 1~GHz. Exploiting this window requires sufficiently rapid alerts and telescope response, particularly for bursts emitted before coalescence. Low-latency GW pipelines are therefore central to multimessenger follow-up, and early-warning alerts with near-zero or negative latency are now a realistic operational goal for at least a subset of events \citep{2020ApJ...905L..25S,2021ApJ...910L..21M,2020ApJ...902L..29N,2022ApJ...927L...9K,2024PNAS..12116474C}. Recent measurements of online-search performance during the fourth observing run (O4) of the Laser Interferometer Gravitational-Wave Observatory (LIGO), Virgo, and KAGRA (collectively, LVK), together with ongoing pipeline-development studies, further characterize current latency and demonstrate paths toward reducing it \citep{2024PNAS..12116474C,2025arXiv251104730H,2026arXiv260425116K,2026arXiv260519153A}. As the LVK GW detector network \citep[e.g.,][]{2015CQGra..32g4001L,2015CQGra..32b4001A,2021PTEP.2021eA101A} improves in sensitivity and localization capability over future observing runs \citep{2020LRR....23....3A}, successful prompt radio follow-up will depend not only on receiving alerts quickly, but also on deciding how to use the first seconds to minutes after an alert most effectively.

Several facilities have explored or proposed prompt radio follow-up of GW events across the low-frequency and GHz radio regimes. Low-frequency arrays such as the Murchison Widefield Array (MWA), the Low Frequency Array (LOFAR), and the Owens Valley Radio Observatory Long Wavelength Array (OVRO-LWA) benefit from very large instantaneous fields of view and, for sufficiently large dispersion measures, dispersive delays that can make post-alert observations sensitive to emission produced near merger time \citep[e.g.,][]{2016PASA...33...50K,2019MNRAS.489L..75J,2019MNRAS.489.3316R,2022MNRAS.509.5018G,2023MNRAS.523.4748G,2023PASA...40...50T,2019ApJ...877L..39C,2025ApJ...985..265K,2026ApJ...997..311K}. However, they are typically less sensitive than higher-frequency facilities. Similar considerations have motivated prompt-burst strategies with the Australian Square Kilometre Array Pathfinder (ASKAP), whose phased-array feeds and fly's-eye observing modes provide very large sky coverage for a GHz-frequency facility, but whose observing band still leaves much shorter dispersive delays than at low radio frequencies, making negative-latency alerts or extremely rapid response especially important \citep{2020PASA...37...51W}. Sensitive GHz arrays therefore occupy a complementary regime: they can reach fainter source limits, but their smaller instantaneous sky coverage makes the allocation of observing time and telescope pointings a central part of the follow-up problem.

This challenge is especially important for the Deep Synoptic Array\footnote{\url{https://www.deepsynoptic.org}} (DSA; formerly DSA-2000), a planned GHz survey and transient-discovery array optimized for real-time radio imaging and commensal fast-transient searches \citep{2019BAAS...51g.255H}. The DSA reference design comprises 1650 steerable 6.15-m dishes distributed over a $\sim20\,{\rm km}\times16\,{\rm km}$ area in a radio-quiet valley in Nevada, with instantaneous frequency coverage from 0.7--2\,GHz, a primary-beam full width half maximum (FWHM) of $\sim2.5^\circ$ at 1.35\,GHz, and a synthesized beam of $\sim3''$ at 1.35\,GHz. Operationally, the DSA antennas share a common commanded pointing; the array is not divided into independently slewing subarrays that can tile multiple localization regions at the same time. Prompt GW follow-up with the DSA is therefore a sequential decision problem: the array must choose a single pointing at each time, while accounting for source observability, finite field of view, slew overheads, beam layout, and the time-dependent probability distribution supplied by the GW alert stream. Although the DSA is also expected to be powerful for longer-timescale radio afterglow studies of compact-binary mergers \citep{2021MNRAS.505.2647D,2024MNRAS.533..510M}, here we focus on the distinct prompt-search problem.

This motivates an adaptive optimization framework for rapid follow-up under uncertainty. Reinforcement learning (RL) is a machine-learning approach in which an agent learns a decision-making policy by interacting with an environment, taking actions based on the current state, and receiving rewards that quantify how well those actions achieve a specified objective. Rather than prescribing the optimal action explicitly, the policy is learned through repeated simulated experience. In the DSA follow-up problem, the telescope acts as the agent, the evolving GW localization and telescope state provide the information on which pointing decisions are based, and the reward measures how effectively the pointing strategy captures localization probability during the prompt observing window. An RL policy can learn when it is worthwhile to slew, when it is better to remain on a high-probability field, and how to balance sky probability against observability and latency without imposing a fixed hand-designed tiling rule. This is attractive because the optimal action depends on the history of previous pointings as well as the current alert information and remaining observing time. RL is already being explored in astronomy for related control and scheduling problems, including adaptive optics control and real-time telescope scheduling, making it a timely approach for transient follow-up optimization \citep{2024A&C....4800833Y,2022A&A...664A..71N,2024AJ....168..214Z,2025arXiv250211134Z}.

In this work, we investigate strategies for optimizing prompt radio follow-up of GW events. In \autoref{sec:optimization}, we develop an RL-based strategy for rapid DSA follow-up of GW candidates, compare it to a baseline strategy, and suggest possible deployment scenarios. In \autoref{sec:test}, we use prototype DSA antennas at the Owens Valley Radio Observatory (OVRO) and present measurements assessing whether they can support such optimized follow-up strategies. In \autoref{sec:phys}, we discuss the sensitivity of the DSA to prompt radio emission from compact-object mergers, the resulting implications for physical models, and the potential for serendipitous gamma-ray burst studies. In \autoref{sec:summary}, we summarize our results.

\section{Observing-strategy optimization}
\label{sec:optimization}

Our framework converts a sequence of GW localization maps into telescope pointing commands. We first describe the simulated alerts and telescope constraints, then define the information supplied to the controller, its pointing actions, and the reward used for training. Training consists of repeated simulated follow-up episodes in which the policy is adjusted to improve probability capture.

\subsection{Simulated GW localization sequences}
\label{sec:simulated_events}

\begin{figure*}[t]
\centering
\includegraphics[width=0.94\textwidth]{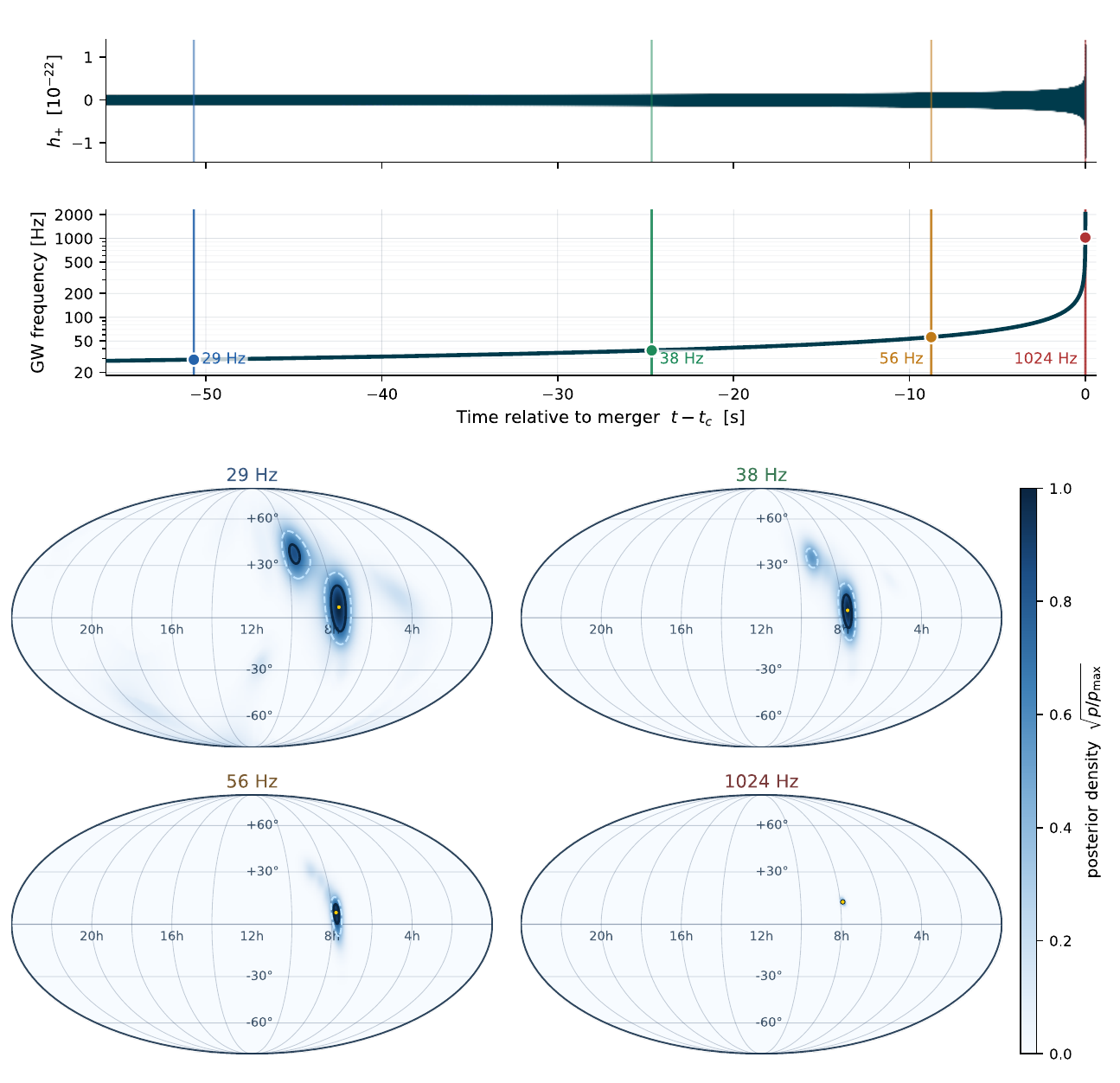}
\caption{ Representative simulated BNS localization sequence used in the follow-up environment. The injection has component masses $1.28+1.83\,M_\odot$, a luminosity distance of $D_L\simeq93$ Mpc, and an inclination of $i\simeq107^\circ$, where $i$ is the angle between the orbital angular momentum and the line of sight. The top panel shows the corresponding plus-polarization strain $h_+(t)$, and the middle panel shows the instantaneous GW frequency, with markers indicating where the signal reaches the BAYESTAR cutoff frequencies of 29, 38, 56, and 1024~Hz (approximately 51, 25, and 9~s before merger, and at merger). The lower panels show the BAYESTAR sky maps generated at each cutoff, which evolve from broad and multimodal at early times to compact near merger; the dot marks the posterior maximum. }
\label{fig:simulated_map_evolution}
\end{figure*}

To train and evaluate the pointing policy, we first constructed a dataset of simulated GW localization sequences for compact-binary mergers involving at least one neutron star. The simulations are intended to represent an O5-like three-detector network, where O5 denotes the planned fifth LVK observing run. The preceding O4 ran from 2023 May 24 to 2025 November 18. LVK observing-scenario studies model future detector sensitivities and expected localization performance for O4/O5-like networks \citep{2020LRR....23....3A,2023ApJ...958..158K}. Current International Gravitational-Wave Observatory Network (IGWN) public-alert documentation treats O5 as a set of plausible sensitivity scenarios, while the official LVK observing-plan page\footnote{\url{https://observing.docs.ligo.org/plan/}} notes that a six-month Intermediate Run 1 (IR1) is planned to begin between late October and mid-November 2026, with O5 plans and the timeline remaining under discussion. We therefore adopt a representative Hanford--Livingston--Virgo (HLV) network, using the LIGO detectors at Hanford, Washington, and Livingston, Louisiana, together with Virgo, with the two LIGO detectors at the target design sensitivity of the Advanced LIGO Plus (A+) upgrade and Virgo at an O5-high sensitivity configuration, as a controlled test case for latency-aware radio follow-up. KAGRA is not included in these simulations; the resulting localizations should therefore be interpreted as a representative three-detector HLV test set rather than as predictions for the full O5 LVK network.

The injected population includes BNS and NSBH systems generated with LALSuite, the LVK Algorithm Library software suite \citep[][]{lalsuite}. BNS component masses were drawn between $1.0$ and $2.3\,M_\odot$. For NSBH systems, neutron-star masses were drawn from the same range, while black-hole masses were drawn between $2.3$ and $5.0\,M_\odot$. These choices are not intended to reproduce the full astrophysical compact-binary population. Instead, they emphasize neutron-star--containing systems most relevant for rapid electromagnetic follow-up: BNS mergers with ordinary to high neutron-star masses, and low-mass NSBH systems for which tidal disruption and matter remaining outside the black hole are more plausible than in high-mass-ratio systems. For NSBH mergers, this outcome also depends sensitively on the black-hole spin and the neutron-star compactness. Therefore, the adopted mass range should be understood as a follow-up-motivated injection choice rather than a guarantee of bright electromagnetic emission \citep[e.g.,][]{2012PhRvD..86l4007F,2018PhRvD..98h1501F}. Component spin magnitudes were limited to $\chi \leq 0.05$, representing a deliberately low-spin injection population appropriate for BNS-like early-warning studies. Sky positions and binary orientations were drawn isotropically. BNS systems were simulated out to $250$ Mpc and NSBH systems out to $400$ Mpc, with multiple distance bins used to populate the training set. These distance limits emphasize nearer and higher signal-to-noise ratio (SNR) events for which early warning and rapid follow-up are most relevant, rather than attempting to reproduce the full O5 public-alert distance distribution.

For all injections, we used the IMRPhenomXPHM waveform approximant as implemented in LALSimulation \citep{2021PhRvD.103j4056P,lalsuite}. Candidate events were required to produce triggers in at least two detectors, with a single-detector SNR threshold of $\rho_i \geq 4$ and a network SNR threshold of $\rho_{\rm net} \geq 12$. This selection is more restrictive than the network-SNR threshold used in LVK public-alert rate forecasts, which are designed to approximate the rate of public alerts rather than to define a high-confidence training sample. We adopt the stricter threshold to select events with robust, self-consistent localization sequences for training and evaluating the follow-up policy.

For each injected event, we generated a sequence of BAYESTAR localizations at selected GW frequency cutoffs. These cutoffs approximate the information available to a low-latency early-warning pipeline as the inspiral signal accumulates in band. Lower cutoff frequencies correspond to earlier pre-merger times and include only the low-frequency portion of the inspiral, whereas higher cutoffs incorporate more signal information. In the simulations used here, the sequence includes early-warning maps at cutoff frequencies of 29, 38, and 56 Hz, together with a final near-merger localization generated using a 1024 Hz cutoff. These values are representative of the frequency-truncated searches used in early-warning studies, where the cutoff frequency acts as a source-dependent proxy for time before merger \citep{2020ApJ...905L..25S,2021ApJ...910L..21M,2020ApJ...902L..29N}; future work can extend the latency-aware DSA pointing framework to other cutoff frequencies and update times. At early times, the limited signal bandwidth and lower effective SNR provide weaker relative timing, amplitude, and phase constraints across the detector network. The resulting localizations are therefore often dominated by timing triangulation and can take the form of broad rings, arcs, or multimodal and mirror-degenerate structures. The prevalence of these structures also depends on the number of detectors contributing to the localization: with all three detectors participating, nearly complete rings are uncommon even early in the inspiral, whereas effectively two-detector cases, including events for which Virgo initially contributes only weak constraints, produce the more extended ring- and arc-like geometries seen in some of our sequences. As higher-frequency inspiral information accumulates, these constraints strengthen and the posterior generally contracts and simplifies \citep{2009NJPh...11l3006F,2011CQGra..28j5021F,2014PhRvD..89d2004G,2015ApJ...804..114B,2018CQGra..35j5002F,2020LRR....23....3A}.

For each cutoff frequency, we generated simulated coincidence data and computed a corresponding Hierarchical Equal Area isoLatitude Pixelation (HEALPix) posterior sky map with BAYESTAR. BAYESTAR provides rapid Bayesian sky localization for compact-binary candidates and is therefore well matched to the low-latency follow-up problem considered here \citep{2016PhRvD..93b4013S,2005ApJ...622..759G}. The coincidences were intially simulated in the zero-noise approximation. As a consequence, each localization sequence represents the noise-free expectation for its injection. The resulting maps should therefore be interpreted as optimistic localizations relative to real low-latency alerts, where measurement noise can broaden posteriors and increase multimodality, especially at low SNR. To test the framework under measurement noise, we additionally regenerate the localization sequences with detector-level Gaussian noise, drawing the matched-filter timing, phase, and amplitude estimates with errors set by the O5 sensitivity curves \citep{2014ApJ...795..105S,2016PhRvD..93b4013S}. Within each event, a single noise realization is shared across all four frequency cutoffs, mirroring a real early-warning analysis in which successive localizations are produced from the same underlying data stream as the signal accumulates \citep{2020ApJ...905L..25S}.

We required each simulated event to have a complete localization sequence across all selected frequency cutoffs. If map generation failed, or if the event did not satisfy the detection criteria at any stage, the event was discarded. This selection ensures that every RL episode contains a self-consistent time sequence of sky maps for the same physical merger. The training set comprised more than ten thousand such localization sequences.

\autoref{fig:simulated_map_evolution} shows one representative localization sequence from the simulated BNS sample. As the frequency cutoff increases, the available inspiral SNR grows and the BAYESTAR posterior contracts from a broad, multimodal early-warning map to a sharper near-merger localization.

\subsection{Environment and observability constraints}
\label{sec:environment}

We model the telescope as a finite-field-of-view (FoV), fast-slewing instrument with realistic observability constraints. Because the DSA antennas share a single commanded array pointing, each action corresponds to a repointing of the full array rather than to assigning independently slewing subarrays to different sky regions. Pointings are required remain above a minimum elevation of \(20^\circ\). The maximum slew speed is set to \(1.5^\circ\,{\rm s}^{-1}\), comparable to the expected array response. Thus, the telescope cannot instantaneously move between widely separated localization modes. Each episode spans $-60$ s to $+20$ s relative to merger and is discretized into 1 s control steps, terminating at $+20$ s after $T=80$ steps regardless of the telescope pointing. This window focuses the optimization on acquiring and tracking the localization before, during, and shortly after coalescence. In the adopted simulation, the final localization in the sequence is assumed to become available near merger and remains the active map for the rest of the episode. The observing horizon could be extended in future work to investigate later emission and incorporate additional localization updates. To sample different visibility geometries, training episodes are initialized at random local sidereal times. We assume that the array is continuously available for GW follow-up throughout the episode. We do not include separate integration, settling, readout, or scheduling overheads, and data loss during slewing is not explicitly modeled. We approximate the DSA field of view as a circular top-hat region with an area of $18\,{\rm deg}^2$.

\begin{figure*}[t]
\centering
\includegraphics[width=\textwidth]{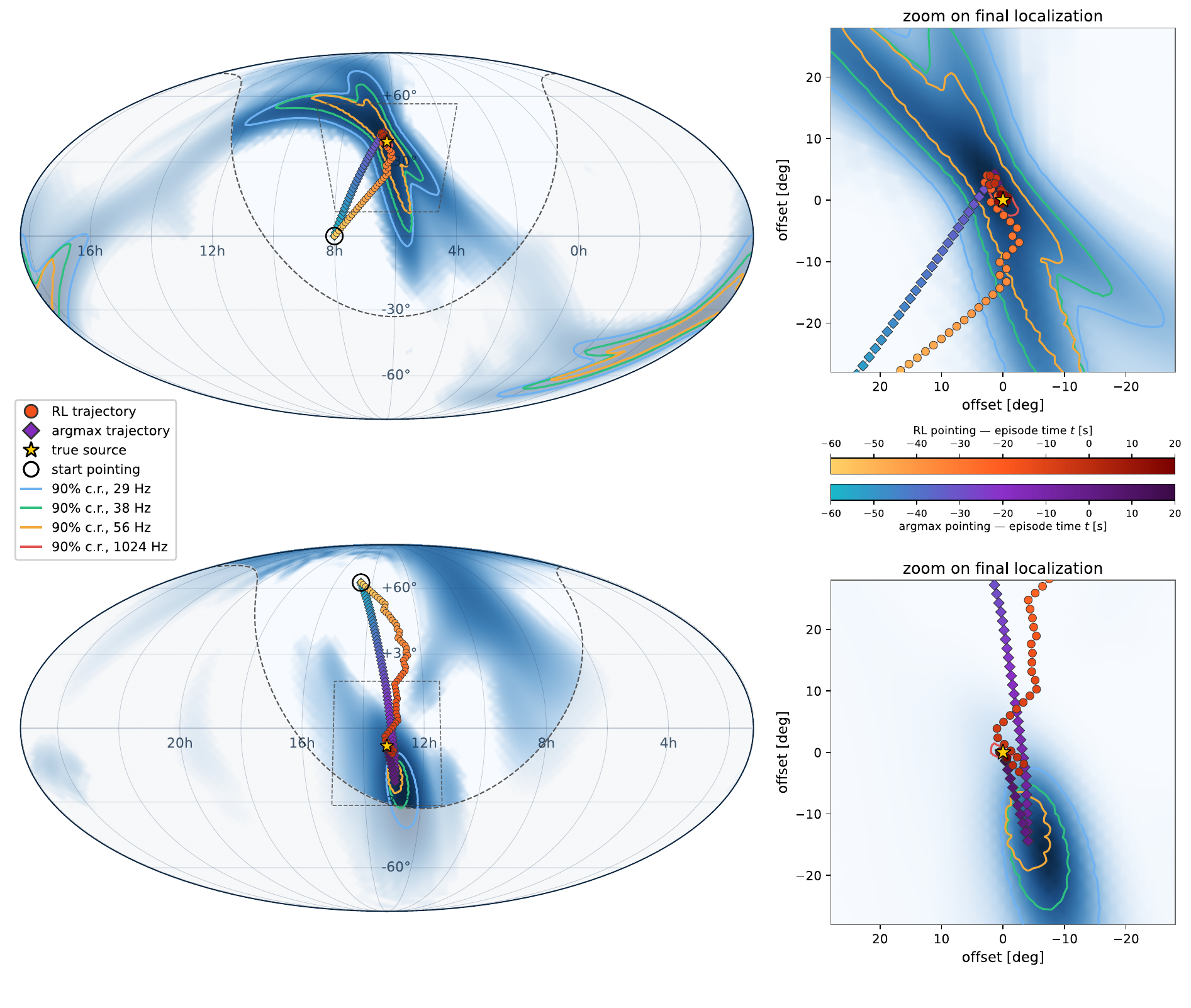}
\caption{
Example paired rollouts comparing the RL policy with the visibility-aware peak-tracking baseline. In each row, the left panel shows the earliest BAYESTAR map (logarithmic stretch) with the $90\%$ credible contours of all four localization stages, the dashed curve marks the $20^\circ$ minimum-elevation boundary, with the inaccessible sky below it shaded.  The pointing trajectories of the two controllers are colored by episode time, and the right panel zooms in on the final localization. Top row: an arc-like localization in which both controllers reach the dominant mode, and the RL trajectory remains slightly better aligned with the broader posterior structure, yielding a capture reward a few percent above the baseline. Bottom row: an event whose pre-merger posterior maxima are noticeably offset from the eventual source position; the peak-tracking controller follows these offset maxima and is still approaching the source at merger, while the learned policy settles onto it earlier and accumulates roughly $40\%$ more capture reward, mostly in the heavily weighted post-merger window.
}
\label{fig:rl_winning_example}
\end{figure*}

\subsection{RL formulation of the pointing problem}
\label{sec:rl_formulation}

We formulate the DSA follow-up problem as a sequential decision problem over a fixed observing window, in which the telescope chooses a new pointing at each step. The simulator maintains the selected localization sequence and its release times, episode time, observing geometry, telescope pointing, accumulated coverage, and the history of actions and pointings.

The policy receives an observation $o_t$ containing all BAYESTAR localization maps released up to the current time, including earlier localization stages but excluding future updates, together with the telescope pointing, visible sky region, accumulated coverage, and compact timing and geometric features. Accumulated coverage records how often each sky pixel has fallen within the simulated telescope field of view during the episode. 

The action is a continuous pointing increment in the local tangent plane, limited by the maximum slew rate. It is converted into the next telescope position; the antenna control software handles the motor motion. The policy therefore has to account for both the evolving probability structure of the GW sky localization and the finite time required to move between separated localization regions.

The policy $\pi(\mathbf{a}_t \mid o_t)$ specifies a distribution of pointing actions given the current observation. The objective is to learn a policy that maximizes the expected cumulative reward over the finite observing window:
\begin{equation}
\pi^\star
=
\arg\max_{\pi}
\mathbb{E}_{\pi}
\left[
\sum_{t=0}^{T-1} r_t
\right],
\end{equation}
where \(T\) is the number of steps in the episode, \(r_t\) is the reward obtained at step \(t\), and the expectation is taken over the simulated event population, observing geometry, initial telescope pointing, and the pointing actions selected by the policy.

The primary reward is the time-weighted localization probability mass inside the telescope field of view,
\begin{equation}
r_t^{\rm cap}
=
\Delta t\, w(t)\, P_{\rm FoV}^{\rm vis}(t),
\label{eq:capture_reward}
\end{equation}
where $\Delta t$ is the control timestep, $P_{\rm FoV}^{\rm vis}(t)$ is the BAYESTAR posterior probability contained within the DSA field of view and satisfying the elevation and declination constraints, and $w(t)$ is a time-dependent weighting factor. The factor $w(t)$ increases toward merger and gives larger weight to the post-merger interval, when prompt electromagnetic emission is most relevant and the policy has access to the most compact localization available in the sequence. Given the uncertainty in the burst arrival time, we optimize probability capture over the observing window rather than at a single epoch. The adopted time weighting emphasizes the interval around and shortly after merger without assuming a specific burst arrival-time distribution. During training we also include potential-based reward shaping toward the dominant visible probability peak and a weak action-smoothness penalty; both are subdominant to the capture reward and are annealed during training.

We train the pointing policy with proximal policy optimization (PPO; \citealt{2017arXiv170706347S}) using the Stable-Baselines3 implementation \citep{JMLR:v22:20-1364} and a custom multi-input feature extractor. The policy input consists of three physically motivated components: a projected sky image encoding the available GW localizations, current pointing, accumulated coverage, and a visibility mask identifying sky pixels that satisfy the elevation and declination constraints at the current time; a vector of scalar quantities such as time to merger, peak offsets, current FoV probability, and visible probability fraction; and a short history of recent actions and pointings. These three input components are summarized in \autoref{tab:obs_config}. Peak offsets describe the direction and angular separation from the current pointing to a probability maximum. The current FoV probability is the sum of localization probability within the visible part of the field of view, while the visible probability fraction is the sum over the entire accessible sky. These inputs allow the network to combine map morphology, the current pointing and visibility constraints, and recent motion. The peak offsets explicitly provide information about the direction of the peak-tracking target. The inputs are processed by a convolutional map encoder, a temporal history encoder, and a shared actor--critic network. Further details of the observation representation, network architecture, reward terms, and training setup are given in \autoref{appendix:rl_details}. We adopted PPO for the initial implementation of the pointing policy. This choice was not based on a comparative evaluation of RL algorithms. Off-policy alternatives, including soft actor--critic (SAC; \citealt{2018arXiv180101290H}), may offer different performance or sample efficiency and will be considered in future work.

\begin{deluxetable*}{lll}
\tablecaption{Observation components supplied to the policy. Further details are provided in \autoref{sec:rl_formulation}
and \autoref{appendix:rl_details}.
\label{tab:obs_config}}
\tablewidth{0pt}
\tablehead{
\colhead{Component} & \colhead{Shape} & \colhead{Content}
}
\startdata
Map tensor & \(7\times64\times64\) &
Four GW localization channels (unreleased stages zero-filled), \\
& & current pointing, accumulated coverage, visibility mask \\
Vector features & 18 &
Timing and pointing features, peak offsets, FoV probability, \\
& & map stage, recent slew magnitude, visibility summaries \\
History buffer & \(5\times80\) &
Two action components and the three Cartesian components \\
& & of the pointing unit vector per step, with capacity for 80 steps \\
\enddata
\end{deluxetable*}

The learned policy is evaluated against a visibility-aware baseline that always slews toward the highest-probability currently visible sky pixel (hereafter the peak-tracking or argmax baseline). Here, argmax denotes the location of this maximum. This baseline is intentionally strong for instantaneous probability capture, but it has no lookahead over future localization updates.

\begin{figure*}[t]
\centering
\includegraphics[width=0.91\textwidth]{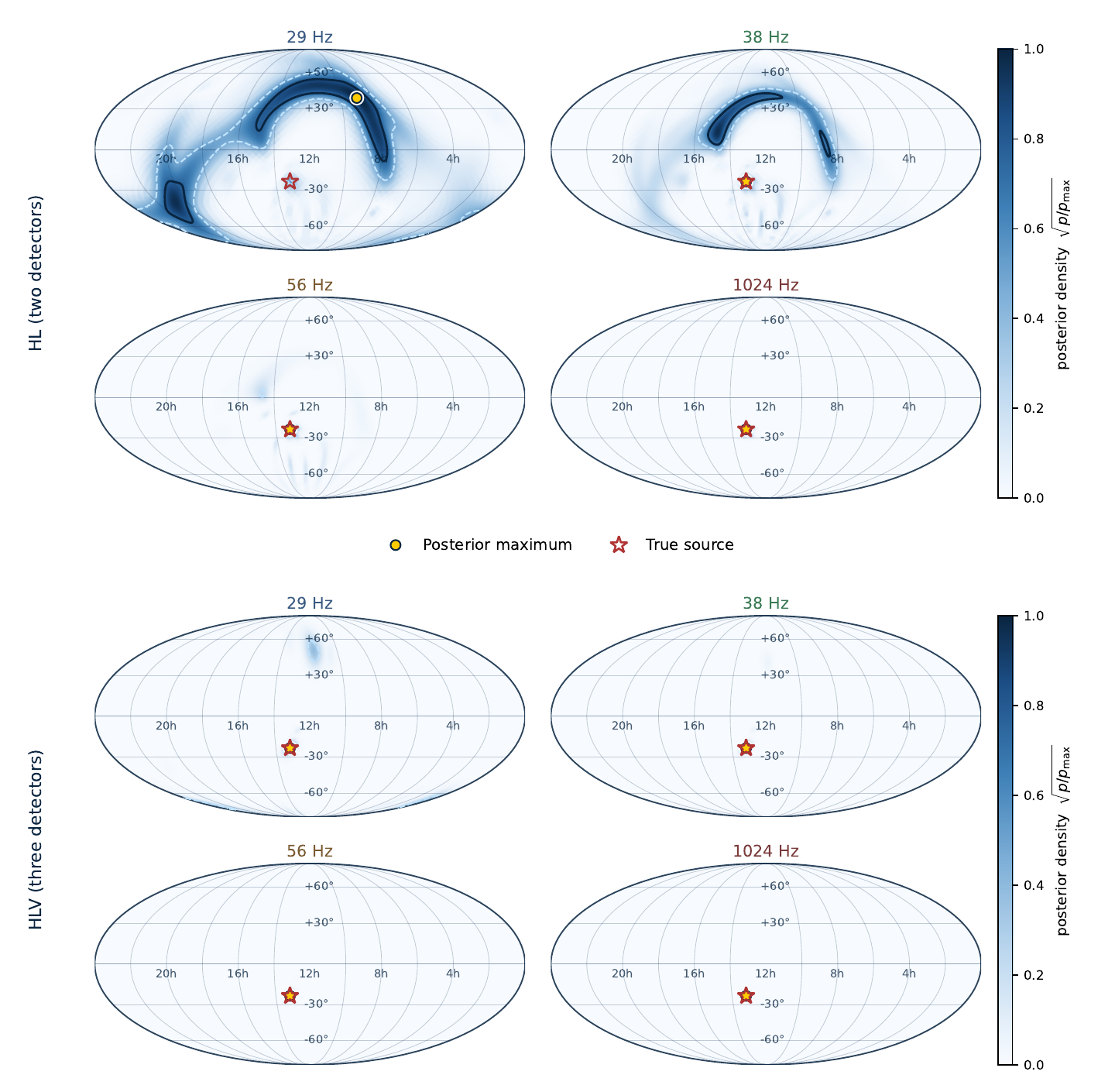}
\caption{GW170817-like localization evolution for a two-detector Hanford--Livingston network with Virgo removed (top block) and a three-detector Hanford--Livingston--Virgo network (bottom block), for the same injection and localization pipeline. Panels show the BAYESTAR sky maps at the 29, 38, 56, and 1024~Hz frequency cutoffs. The star marks the true source position and the dot the posterior maximum. With three detectors the maximum coincides with the true position from the earliest map, whereas the two-detector posterior is an extended arc whose maximum lies tens of degrees from the source until shortly before merger.
}
\label{fig:gw170817_evolution_hlv_hl}
\end{figure*}

\subsection{Main results of optimization}
\label{sec:optimization_results}
The visibility-aware peak-tracking baseline provides a strong reference strategy for prompt DSA follow-up. In nominal localization sequences, where the posterior is compact or dominated by a single visible mode, the peak-tracking strategy is difficult to beat because it directly targets the highest-probability accessible sky pixel. The trained RL policy achieves comparable cumulative probability capture in these cases, indicating that it learns the basic structure of the pointing problem. Pointing at the posterior maximum is optimal only for a point-like target with unconstrained slewing; with a finite field of view, slew limits, and evolving maps, the optimal pointing maximizes the beam-enclosed probability over the observing window and can therefore differ from instantaneous peak tracking.

The main benefit of the learned policy appears in less ideal localization regimes. For broad, multimodal, or arc-like posteriors, finite slew time makes the instantaneous posterior maximum an incomplete description of the best observing action. A small subset of test cases with these localization morphologies can favor the RL policy, although broadness alone is not sufficient to identify them. We therefore explore a lightweight random-forest
classifier for selecting between the two controllers. Using summary features of the earliest localization map, the classifier predicts whether the learned policy is likely to outperform the peak-tracking controller, allowing the choice to be made before merger. On the exploratory held-out set, a conservative threshold of \(P_{\rm win}\geq0.65\) routes approximately \(2\%\) of paired test cases to RL; within this routed subset, the summed post-merger capture score is \(3.15\%\) higher than for peak tracking, although RL does not win every individual comparison (see \autoref{appendix:rl_details}).

\autoref{fig:rl_winning_example} shows two representative paired rollouts in which the RL policy outperforms the visibility-aware peak-tracking baseline. The examples illustrate cases in which the learned policy uses the evolving localization sequence more effectively than a controller that follows only the instantaneous posterior maximum.

The learned policy performs comparably to peak tracking in most paired simulations, indicating that it recovers the behavior of the simpler controller in nominal conditions. Peak tracking is nevertheless attractive for initial deployment because it follows a deterministic rule for a given localization map and telescope state, its decisions can be inspected directly, and its implementation is straightforward to validate and maintain. The classifier explores whether cases offering an additional benefit from RL can be identified in advance; it is not required by the RL formulation. The learned policy can also be used directly for all alerts without a classifier. Whether selective deployment justifies the additional complexity of maintaining two controllers and a selector remains to be established.

\autoref{fig:gw170817_evolution_hlv_hl} shows a GW170817-like localization sequence simulated with the same pipeline for the full HLV network and for a two-detector Hanford--Livingston network with Virgo removed. With three detectors, the source is localized to a compact region already in the earliest map. With two detectors, timing triangulation constrains the source only to an extended arc that remains broad through merger. Two-detector alerts of this type are precisely the regime in which instantaneous peak-tracking is least informative about where to point and where a history-aware policy has the most room to help. The present comparisons use zero-noise localization sequences, but real alerts may be less well localized because of measurement noise or reduced detector participation. To test how the framework behaves under measurement noise, we fine-tuned the policy on the Gaussian-noise sequences of \autoref{sec:simulated_events} and repeated the paired comparison (see more details in \autoref{appendix:rl_details}). The picture is similar to the noiseless case: the two controllers remain comparable for most alerts, and a targeted subset of localizations retains an RL advantage on new noise realizations not used during training, but routing under noise requires a more sophisticated dispatch model --- a regression on the same five early-map features rather than a binary classifier. More realistic, non-Gaussian noise models remain a direction for future work.

\subsection{Follow-up reachability for the simulated population}
\label{sec:reachability}

The preceding analysis quantifies how well a given localization can be tracked. We also estimate what fraction of the simulated population could be followed up during the prompt window. For this purpose, we evaluated the controller on the simulated O5 HLV population starting from a random initial pointing on the accessible sky, using each event's chirp-mass-dependent early-warning times and the antenna slew rates measured in \autoref{sec:test_results}. This estimate is intended only as an order-of-magnitude assessment of geometric reachability, rather than a full observing-yield calculation.

At a random merger time, $37\%$ of the simulated events are above the adopted $20^\circ$ minimum elevation and are therefore immediately observable. Among this observable subset, BNS systems are reachable from a random initial pointing in $\sim90\%$ of cases, with the source acquired and kept within the primary beam through the prompt window. The reachable fraction is lower for NSBH systems: using the measured slew rates, it is $\sim$55\% for lower-mass NSBH systems and can fall to $\sim$35\% for higher-mass systems. This difference is driven by the shorter warning time for NSBH mergers, for which the first localization arrives only $\sim$15--20\,s before merger; in this case, the time required to slew from the initial pointing can become the limiting factor. 

Repeating the acquisition calculation on sequences regenerated with Gaussian measurement noise lowers these conditional fractions: the BNS acquisition fraction decreases from $\sim$90\% to $\sim$75\%, and the corresponding NSBH warning-time proxies decrease from $\sim$55\% to $\sim$47\% (lower-mass) and from $\sim$35\% to $\sim$30\% (higher-mass), as the noisy early maps point the initial slew less reliably toward the true source. The instantaneous observability fraction of $37\%$ is set by source position and elevation geometry and is unaffected by map noise. These estimates remain approximate: the calculation does not include the real distribution of alert latencies and localization areas, and a more realistic calculation of the follow-up yield is left to future work.

\section{Validation with prototype DSA antennas}
\label{sec:test}

\begin{figure*}[t]
\centering
\includegraphics[width=\textwidth]{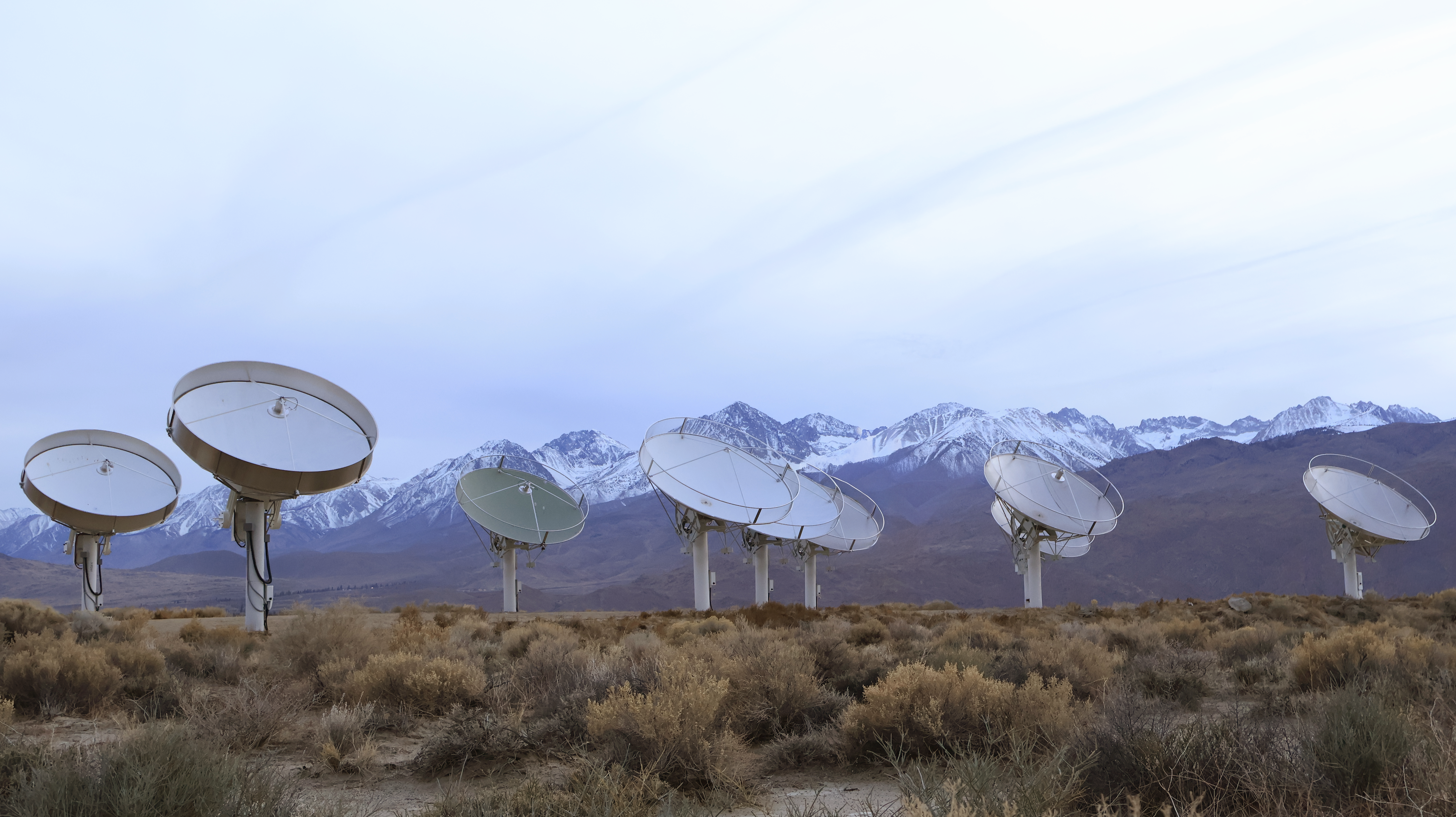}
\caption{Prototype Deep Synoptic Array (DSA) antennas at the Owens Valley Radio Observatory. The test array includes multiple prototype antenna generations developed as part of the DSA program. The two leftmost antennas in this image were used for the trajectory-playback tests presented in this work. These tests provide a pathfinder validation of rapid follow-up operations, including antenna repointing, slew-time constraints, and the interface between optimized pointing policies and real telescope control. Credit: Vinand Prayag.}
\label{fig:test_array}
\end{figure*}

\subsection{Antenna control interface and trajectory playback}
\label{sec:test_software}

In this section, we use two prototype DSA antennas at the Owens Valley Radio Observatory to test the execution of optimized pointing trajectories with real antenna hardware. The broader DSA antenna program has proceeded through multiple generations of design and development. The original 5-m-aperture design is described by \citet{2025usnc.conf..319W}, with a subsequent 5-m design variation led by mtex Antenna Technology GmbH.\footnote{\url{https://www.mtex-at.com}} The current reference design for the array has a 6.15-m aperture and is an evolution of the mtex 5-m design. The trajectory-playback tests were performed using the two leftmost antennas shown in \autoref{fig:test_array}, which are examples of the mtex 5-m design variant. The discussion below refers specifically to the two antennas used in these tests and their control interface.

The prototype antennas are commanded through the DSA monitor-and-control software stack, in which an array-controller library sends commands over a message-bus interface to the antenna control units. Each antenna exposes both direct position commands and a programmed-track mode, in which the control unit follows a list of time-tagged azimuth--elevation samples anchored to an absolute start epoch, interpolating linearly between samples.

The successive policy actions generate a sky-pointing
trajectory, which is converted into time-tagged azimuth
and elevation coordinates for the programmed-track
interface. To connect the pointing policies of \autoref{sec:optimization} to this system, we contributed a thin azimuth--elevation tracking layer to the DSA control software.\footnote{\url{https://gitlab.com/dsa-2000/dsa-2000-repos/dsa-2000-monorepo/}} The playback tool validates each policy trajectory against the antenna drive limits, slews the antenna to the first sample, and then starts the programmed track at the specified absolute epoch.

In these trajectory-playback tests, we neglect the time required to receive and process a GW alert; instead, we test the execution of policy trajectories that have already been generated. For the tests, we used several representative trajectories, with durations and commanded slew rates comparable to the adaptive retargeting expected during a real early-warning alert.

\subsection{Results}
\label{sec:test_results}

We executed multiple trajectories on two prototype antennas on 2026 June 12, with wind speeds of $\sim$8\,m\,s$^{-1}$. All runs completed without faults. The DSA monitoring system records drive telemetry, including encoder positions and velocities at up to 10\,Hz and the servo controller's internal tracking error, which we use for the analysis below.

\begin{figure}[t]
\centering
\includegraphics[width=1\columnwidth]{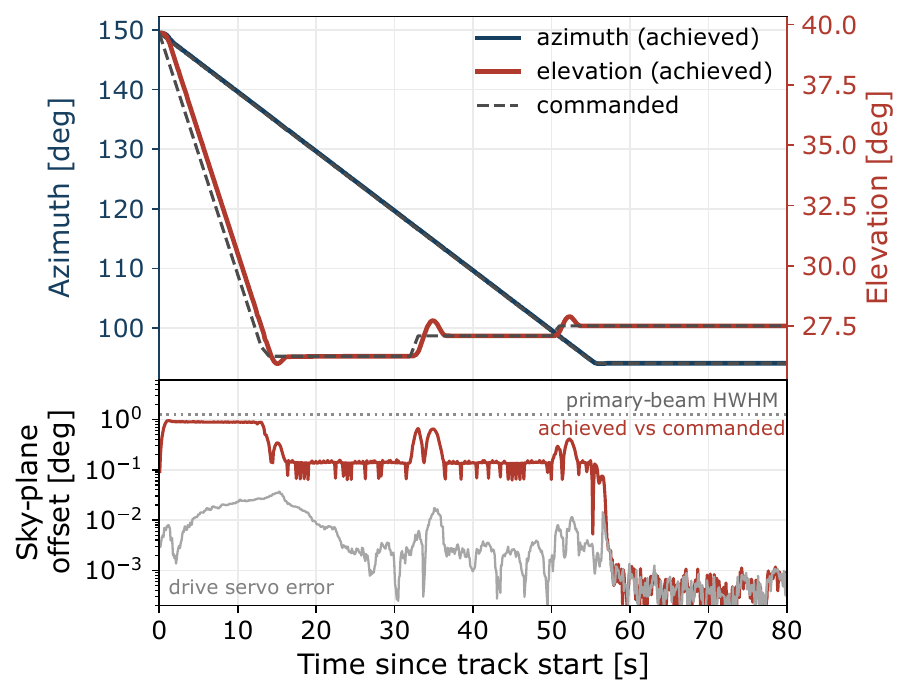}
\caption{Prototype DSA antenna executing one of the simulated GW follow-up trajectories through the programmed-track interface. Top: commanded (dashed) and achieved (solid, encoder telemetry) azimuth and elevation. The commanded curve is the precomputed trajectory supplied to the antenna controller. Bottom: total sky-plane offset between the achieved pointing and the requested trajectory (red), and the drive's internal servo error, i.e.\ the achieved position relative to the controller's own smoothed setpoint (grey); the dotted line marks the primary-beam half width at half maximum (HWHM, $1.25\degr$). The difference between the two curves is due to the drive's internal trajectory shaping rather than to servo-following error: the commanded ramps are delayed by $\sim$0.9\,s on the elevation axis and $\sim$0.1\,s in azimuth, producing the initial plateau while both axes ramp, the $\sim$$0.1\degr$ floor during the remaining azimuth ramp, and the brief excursions at the elevation steps. The curves coincide once the commanded rates drop to zero.}
\label{fig:track_traj03}
\end{figure}

The prototype antennas executed the commanded trajectories, as illustrated by one example in \autoref{fig:track_traj03}. The achieved pointing lags the requested trajectory by up to $\simeq$1\,s during the fastest commanded ramps --- corresponding to offsets of up to $\sim$$1\degr$ in the initial part of the track in \autoref{fig:track_traj03} --- but converges to well within the primary beam on the remainder of the trajectory, with servo errors at the arcsecond level throughout. The repointing slews between runs reached $3.8\degr\,{\rm s}^{-1}$ in azimuth, exceeding the $1.5\degr\,{\rm s}^{-1}$ adopted in \autoref{sec:environment}, while elevation slews ran near $1\degr\,{\rm s}^{-1}$; a fuller characterization of the achievable slew envelope is beyond this work and will be addressed in future tests. Overall, these runs demonstrate that the prototype antennas can execute dynamic, pre-planned pointing trajectories and slew rapidly enough to reach an evolving localization on the timescales relevant for early-warning follow-up.

\section{Physical implications}
\label{sec:phys}

\subsection{Prompt-burst search capabilities}
\label{sec:dsa_sensitivity}

The DSA backend architecture is planned to support multiple real-time graphics processing unit (GPU)-based data streams. The radio-camera stream will produce wide-field image cubes with a projected continuum sensitivity of approximately $600~{\rm nJy}$ in a one hour integration, providing the deep-imaging view of the GHz radio sky. This continuum imaging mode is distinct from the prompt-burst search considered here and does not set the integration time for our follow-up strategy. For prompt coherent emission on millisecond timescales, however, the relevant backend is the DSA Chronoscope. Chronoscope is designed to form dense beam grids across the primary beam for FRB searches. Further details of the Chronoscope backend and its planned FRB search are given by \citet{2026arXiv260830943C}. The FRB-search mode operates at 1 ms time resolution and approximately 130 kHz frequency resolution, with low-latency FRB alerts among the planned data products. Beyond the prompt-burst window, follow-up for longer-lived continuum emission constitutes a separate observing problem. Well-localized events could be monitored with a single pointing, while broader localizations may require mosaicking or tiling; optimization of this longer-timescale afterglow strategy is beyond the scope of this work.

This makes Chronoscope the natural backend for prompt GW follow-up searches for short-duration coherent radio bursts. A simple radiometer estimate gives a characteristic 1 ms sensitivity in the mJy regime, even when using only a fraction of the full DSA bandwidth. For example, using one quarter of the $1.3\,{\rm GHz}$ band and an effective system-equivalent flux density (SEFD) of $1.8\,{\rm Jy}$ gives $\sigma_{\rm 1\,ms}\simeq2\,{\rm mJy}$ before applying a detection-threshold factor. This estimate applies near zenith and represents an optimistic instrumental sensitivity. In practice, sensitivity will degrade away from zenith, and additional losses from the commissioned search configuration, radio-frequency interference (RFI), beam-grid response, and calibration are expected to reduce the effective sensitivity by roughly a factor of two. Poor calibration during rapid follow-up may introduce additional decorrelation on longer baselines. We therefore adopt a representative practical fluence scale of $F_{\nu,\lim}\sim20\,{\rm mJy\,ms}$ for the order-of-magnitude estimates below, while noting that the final threshold will depend on the commissioned system.

The beamformed search can also trade sensitivity against sky coverage. The Chronoscope search beams are produced through a correlate--grid--image processing architecture, while the downstream search operates on a dense grid of synthesized sky beams. At the lower end of the DSA band, the primary beam is broader: at $0.7\,{\rm GHz}$, the FWHM field of view is approximately $18\,{\rm deg}^2$. By allocating beams over a wider footprint, the search can also extend beyond the nominal half-power field of view. For a Gaussian beam approximation, searching to the 20\% primary-beam response level gives an effective area of order $40\,{\rm deg}^2$. The Chronoscope configuration can therefore be tuned between sensitivity and instantaneous sky coverage depending on the localization area and alert priority.

A complementary mode is voltage buffering. For the brightest or highest-priority GW candidates, a voltage buffer retaining approximately one minute of data would preserve the relevant data around the trigger time and allow offline reprocessing with a refined beam grid, updated sky map, or alternative dedispersion/search configuration. The fraction of events for which this mode is used can be chosen adaptively, depending on the real-time alert rate and scientific priority of candidates during future LVK observing runs.

\subsection{Energy reach and GHz-search context}
\label{sec:energy_reach}

For a burst with fluence threshold $F_{\nu,\lim}$ and effective bandwidth $\Delta\nu$, the limiting isotropic-equivalent radio energy is
\begin{equation}
\begin{aligned}
E_{\rm radio,iso}^{\lim}
&\simeq
2.2\times10^{36}
\left(\frac{F_{\nu,\lim}}{20\,{\rm mJy\,ms}}\right)
\left(\frac{\Delta\nu}{1~{\rm GHz}}\right) \\
&\quad \times
\left(\frac{D_L}{300~{\rm Mpc}}\right)^2
{\rm erg}.
\end{aligned}
\label{eq:eiso_numeric}
\end{equation}
For a 1 ms burst, this corresponds to a luminosity scale of $\sim10^{39}\,{\rm erg\,s^{-1}}$ at $D_L=300$ Mpc. Thus, for nearby neutron-star mergers at $D_L\sim100$--$300$ Mpc, the DSA could probe millisecond coherent bursts with radio energies of order $10^{35}$--$10^{36}$ erg and luminosities of order $10^{38}$--$10^{39}$ erg s$^{-1}$.

This distance range is most relevant for the nearby, high-SNR subset of neutron-star-containing events in an O5-like detector network. Current public O5 projections place the BNS range of the LIGO detectors at a few hundred Mpc, with Virgo and KAGRA contributing smaller ranges, although the detailed O5 timeline and achieved sensitivities remain subject to change. We therefore use $100$--$300$ Mpc only as a reference scale, not as a prediction for the full detected population.

GHz prompt-radio searches with other facilities highlight the same basic trade-off. ASKAP studies of prompt emission from neutron-star mergers emphasize that, at GHz frequencies, cold-plasma dispersion delays are short and post-merger alerts are usually too late to capture emission produced at coalescence; negative-latency alerts or very rapid response are therefore essential \citep{2020PASA...37...51W}. ASKAP follow-up of Fermi GRBs at 1.2--1.4 GHz found no associated bursts above fluence limits of order tens of Jy ms, with the strongest interpretation applying to delayed FRB-like emission rather than truly merger-time emission \citep{2020MNRAS.497..125B}. The DSA would instead target much deeper fluence limits over a smaller instantaneous field of view, making optimized pointing central to the search.

\subsection{Interpretation for coherent-emission models}
\label{sec:model_comparison}

The DSA sensitivity scale is best interpreted primarily in a model-agnostic way. Proposed prompt coherent-emission channels include pre-merger magnetospheric interactions, merger-time bursts, delayed collapse of a supramassive remnant, and magnetized post-merger outflows \citep[e.g.,][]{1996A&A...312..937L,2001MNRAS.322..695H, 2013PASJ...65L..12T,2014A&A...562A.137F,2020ApJ...893L...6M, 2023PhRvL.130x5201M}. These models differ strongly in luminosity, beaming, duration, spectrum, and whether the radio emission can escape the dense merger environment. We therefore do not assign the DSA threshold to a single fiducial model.

A radio energy of $10^{35}$--$10^{36}$ erg is a small fraction of plausible magnetic, remnant, or relativistic-outflow energy reservoirs. For example, relative to an available engine energy of $10^{46}$--$10^{49}$ erg, the required isotropic-equivalent coherent-radio fraction is of order $10^{-10}$--$10^{-13}$ at $D_L\sim300$ Mpc. This does not by itself imply that all models are detectable: the relevant quantity is the fraction that emerges as observable GHz coherent emission in the DSA band and along our line of sight.

For context, recent compact-binary magnetosphere simulations provide useful comparison scales for the DSA. For BNS systems, force-free simulations predict precursor flare luminosities of order $L_{\rm EM}\sim10^{44}$--$10^{45}\,{\rm erg\,s^{-1}}$ for surface magnetic fields $B\sim10^{12}$ G; adopting radio efficiencies $\sim10^{-5}$--$10^{-3}$ gives possible coherent-radio luminosities of order $10^{39}$--$10^{42}\,{\rm erg\,s^{-1}}$ \citep{2022MNRAS.515.2710M}. A related reconnection-powered model predicts millisecond radio bursts at characteristic frequencies of $\sim10$--$20$ GHz for $B^\ast\sim10^{11}$ G \citep{2023PhRvL.130x5201M}. For NSBH systems, the predicted electromagnetic luminosities are lower, $L_{\rm EM}\lesssim10^{41}(B_\ast/10^{12}\,{\rm G})^2\,{\rm erg\,s^{-1}}$ \citep{2023ApJ...956L..33M}. Recent magnetohydrodynamic (MHD) simulations of NSBH mergers additionally predict a one-off FRB-like signal from magnetospheric shocks near merger \citep{2025ApJ...982L..54K}; synchrotron-maser emission at such shocks provides a possible GHz radiation mechanism \citep{2025PhRvL.134c5201V}. These values overlap the DSA luminosity scale for nearby events, but the comparison remains order-of-magnitude because the observable signal depends on radio efficiency, beaming, propagation through the merger environment, and whether the coherent spectrum falls within the DSA band. Dense post-merger ejecta can significantly attenuate radio emission and may prevent prompt bursts from escaping in some BNS scenarios \citep{2024ApJ...977..122B,2024MNRAS.528.5836R}. Collapse of a long-lived magnetized remnant may produce a millisecond-peaked high-energy transient \citep{2024ApJ...974L..12M}.

\subsection{Serendipitous overlap with gamma-ray bursts}
\label{sec:grb_overlap}

The Chronoscope FRB-search mode also enables a serendipitous search for prompt coherent radio bursts associated with gamma-ray bursts (GRBs) during normal DSA operations. Previous searches for prompt or early-time radio emission from GRBs include low-frequency campaigns with the MWA and LOFAR, GHz follow-up with ASKAP, and wide-field searches with the Canadian Hydrogen Intensity Mapping Experiment (CHIME)/FRB project \citep[e.g.,][]{2015ApJ...814L..25K,2021PASA...38...26A,2022PASA...39....3T,2022MNRAS.514.2756T,2025ApJ...982...32X,2019MNRAS.490.3483R,2021MNRAS.506.5268R,2023MNRAS.526..106H,2025MNRAS.544...53H,2020MNRAS.497..125B,2023ApJ...954..154C,2024ApJ...972..125C,2024MNRAS.534.2592R}. GRBs are distributed approximately isotropically on the sky \citep{1992Natur.355..143M}, so the probability that a randomly occurring GRB lies within an instantaneous DSA field of view $\Omega_{\rm FOV}$ is simply
\begin{equation}
    f_{\rm sky}
    =
    \frac{\Omega_{\rm FOV}}{4\pi}
    \simeq
    \frac{18}{41253}
    \simeq
    4.4\times10^{-4},
\end{equation}
where we have adopted $\Omega_{\rm FOV}\simeq18~{\rm deg^2}$.

The current all-sky GRB alert rate is of order a few hundred events per year.  For example, the Fermi Gamma-ray Burst Monitor (GBM) detected 2356 cosmic GRBs during its first ten years of operation, corresponding to $\simeq 240~{\rm yr^{-1}}$ \citep{2020ApJ...893...46V}, while the Swift Burst Alert Telescope (BAT) detects of order $\sim 100~{\rm yr^{-1}}$ with arcminute-scale localizations \citep{2005SSRv..120..143B,2016ApJ...829....7L}.  Accounting for overlap between instruments, an order-of-magnitude unique network rate of $R_{\rm GRB}\sim300$--$500~{\rm yr^{-1}}$ is therefore reasonable for estimating accidental overlap.  The expected waiting time for a GRB to fall within the DSA field of view is then
\begin{equation}
    t_{\rm coinc}
    \simeq
    \left(R_{\rm GRB} f_{\rm sky}\right)^{-1}
    \sim
    5\text{--}8~{\rm yr}.
\end{equation}

This estimate is intended only as an approximate serendipitous-overlap rate. The timescale for short GRBs alone is longer because they make up only a minority of the observed GRB population. Nevertheless, short GRBs are especially relevant here because GW170817 and GRB 170817A established that at least some binary neutron star mergers produce short GRBs \citep{2017ApJ...848L..13A,2017ApJ...848L..14G}. Therefore, although blind overlap with a short GRB is rare, a sufficiently bright prompt radio burst associated with such an event could in principle be detected serendipitously within the DSA field of view during normal survey operations.

\section{Summary}
\label{sec:summary}

We have developed and tested a framework for prompt radio follow-up of GW events with the DSA, targeting millisecond-duration coherent emission from neutron-star mergers in the early-warning regime. We constructed simulated low-latency localization sequences for an O5-like HLV network, generating BAYESTAR sky maps at frequency cutoffs of 29, 38, 56, and 1024\,Hz for injected BNS and NSBH events, and embedded them in a follow-up environment with elevation, field-of-view, and slew constraints. Within this environment we trained a reinforcement-learning pointing policy with PPO and compared it with a strong visibility-aware peak-tracking baseline. The two perform comparably for most alerts, indicating that the learned policy recovers the behavior of the peak-tracking rule in nominal conditions.

The learned policy can be preferable for a subset of broad, asymmetric, or multimodal localizations, where a field of view centered on the instantaneous posterior maximum captures only a small fraction of the probability. This regime includes the extended arcs characteristic of two-detector alerts, as illustrated by our GW170817-like sequences with and without Virgo. We explore whether features of the earliest localization and initial telescope pointing can identify cases that benefit from RL. Peak tracking is attractive for initial deployment because of its simplicity and interpretability. The learned policy can be used directly for all alerts, while the classifier provides an optional way to retain peak tracking for most alerts and select RL where an additional benefit is expected.

We further validated the path from optimized pointing sequences to real telescope control. We contributed an azimuth--elevation trajectory-playback capability to the DSA control software and executed pre-planned pointing trajectories on prototype antennas at OVRO. The tests showed that the prototype antennas can execute dynamic, pre-planned pointing trajectories, with the achieved pointing settling. This validates the feasibility of executing such trajectories on timescales relevant for early-warning follow-up.

With the planned Chronoscope beamformed backend, optimized DSA follow-up can reach fluence limits of $\sim$20\,mJy\,ms, probing millisecond bursts with isotropic-equivalent energies of $10^{35}$--$10^{36}$\,erg and luminosities of $10^{38}$--$10^{39}$\,erg\,s$^{-1}$ at 100--300\,Mpc --- at or below the lower end of several coherent-emission predictions for neutron-star mergers. Future work includes evaluation on non-Gaussian, nonstationary detector noise, including injections into real detector data, integration with the real-time LVK alert stream, a fuller characterization of the antenna slew envelope, and deployment of the peak-tracking controller within DSA operations during upcoming observing runs.

%% Please use the acknowledgment and contribution environments. This will 
%% be anonomyized when the "anonymous" style option is used. 
\begin{acknowledgments}
We thank Elijah Malaby for assistance with the DSA prototype antennas and Fabian Walter and Elias R. Most for helpful comments on the manuscript. We also thank the anonymous referee for helpful comments that improved the manuscript. This material is based in part upon work supported by the National Science Foundation under grant number AST-1911199. MM, JK, and AJW were supported by NSF Grant PHY-2309200. This material is based upon work supported by NSF’s LIGO Laboratory, which is a major facility fully funded by the National Science Foundation. LIGO was constructed by the California Institute of Technology and Massachusetts Institute of Technology with funding from the National Science Foundation and operates under cooperative agreement PHY-2309200. This research is also enabled by development funded by Schmidt Sciences for the Deep Synoptic Array project (DSA). The DSA is part of the Eric and Wendy Schmidt Observatory System in partnership with the California Institute of Technology. N.K. acknowledges support from the Schmidt Academy for Software Engineering at the California Institute of Technology. OpenAI ChatGPT and Anthropic Claude were used for limited language editing, code debugging, and figure preparation. All content was reviewed and verified by the authors.
\end{acknowledgments}

\appendix
\section{Reinforcement-learning implementation details}
\label{appendix:rl_details}

This appendix summarizes the RL implementation details: the observation representation, policy architecture, reward terms, training setup, and evaluation.

\subsection{Observation representation}
\label{app:rl_observation}

The policy observation comprises the map channels,
scalar features, and action--pointing history described
in \autoref{sec:rl_formulation}. BAYESTAR HEALPix maps are converted to a common \(N_{\rm side}=32\) resolution and projected onto a \(64\times64\) local sky grid. The projected map channels provide the policy with all BAYESTAR localization maps released up to the current timestep, together with the telescope pointing, accumulated coverage, and a visibility mask identifying sky pixels that satisfy the elevation and declination constraints at that timestep. Future localization stages are zero-filled until their corresponding reveal times, so the input dimensionality remains fixed throughout the episode.

The scalar vector features include the normalized episode time, current pointing unit vector, local-frame offset and angular separation to the dominant visible probability peak, time to the next map update, current FoV probability mass, analogous quantities for a secondary peak, the current map stage, recent slew magnitude, peak altitude, and the total visible probability fraction. The time to the next map update follows the prescribed simulation schedule; its use with live alerts would require an estimate of the next update time.

\subsection{Policy architecture and PPO training}
\label{app:rl_policy}

\begin{table*}
\centering
\caption{Policy-network architecture.}
\label{tab:policy_architecture}
\begin{tabular}{lc}
\hline
Component & Configuration \\
\hline
Map encoder & Conv2D \(7\!\to\!32\), Conv2D \(32\!\to\!64\), Conv2D \(64\!\to\!64\), Linear \(\to512\) \\
History encoder & Conv1D \(5\!\to\!32\), Conv1D \(32\!\to\!64\), adaptive pooling, Linear \(\to64\) \\
Vector branch & 18-dimensional pass-through \\
Concatenated feature size & \(512+18+64=594\) \\
Shared multilayer perceptron & \(594\to256\to256\to128\), with \(\tanh\) activations \\
Actor head & Linear \(128\to2\), diagonal Gaussian action distribution \\
Critic head & Linear \(128\to1\) \\
Action squashing & \(\tanh\) to \([-1,1]^2\), then scaled to physical slew \\
Approximate parameter count & \(1.6\times10^6\) \\
\hline
\end{tabular}
\end{table*}

\begin{figure*}
\centering
\includegraphics[width=0.90\textwidth]{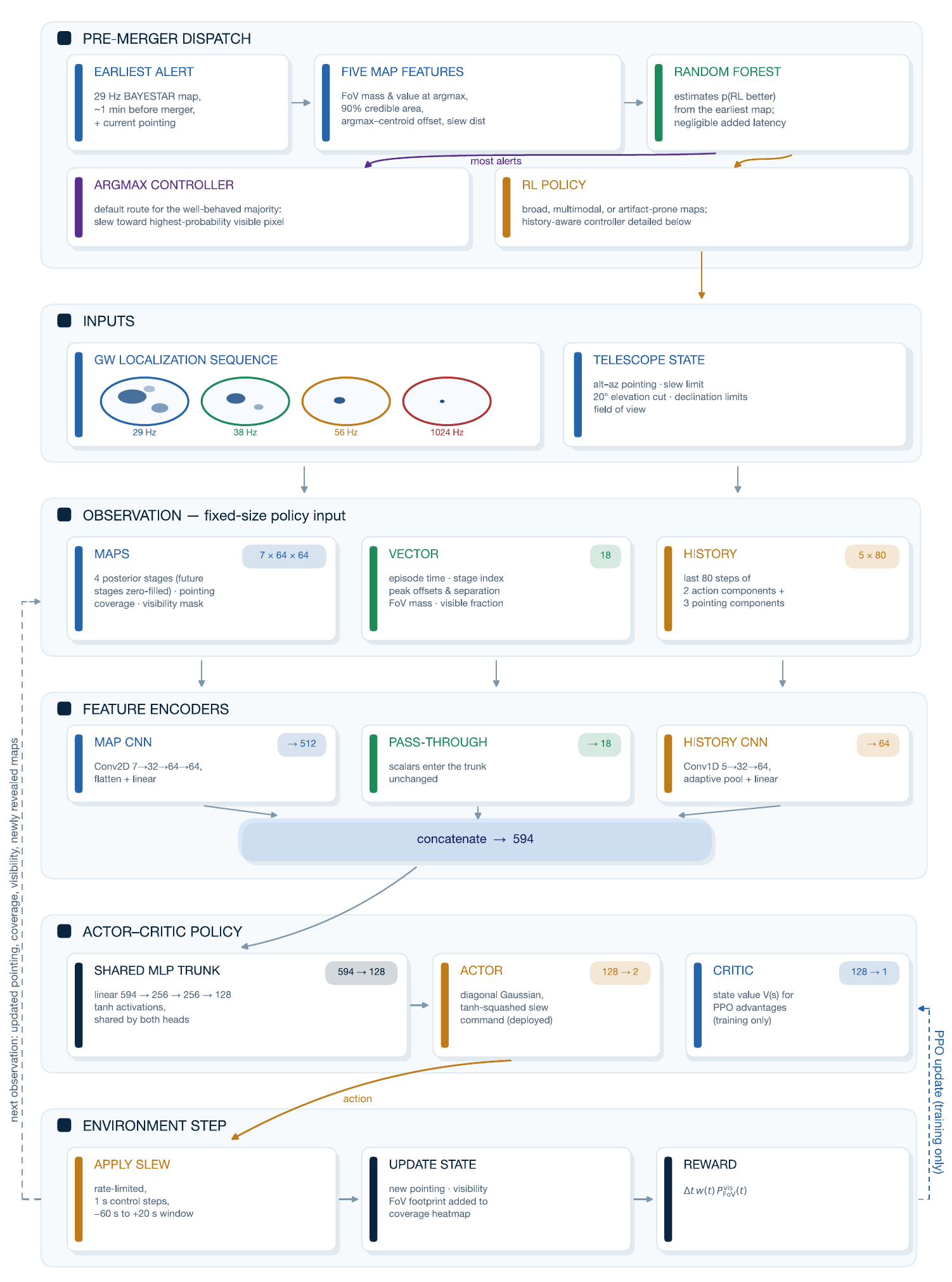}
\caption{Architecture of the explored pointing framework, including an optional classifier-based dispatch stage. Top: the pre-merger dispatch stage summarizes the earliest BAYESTAR map into five features, and a random-forest classifier routes most alerts to the visibility-aware peak-tracking controller, while sending selected alerts predicted to benefit from the learned policy to the RL controller. Bottom: the RL pointing policy. BAYESTAR maps and the telescope state are converted into a fixed observation (projected sky-map channels, scalar features, recent history), processed by separate map, vector, and history encoders, and passed to a shared actor--critic network; the actor outputs a bounded slew command and the environment updates the pointing, visibility, coverage, and reward. The critic and PPO update are used only during training.
}
\label{fig:rl_architecture1}
\end{figure*}

The policy uses a custom feature extractor followed by a PPO actor--critic network, summarized in \autoref{tab:policy_architecture} and \autoref{fig:rl_architecture1}. The map tensor is processed by a two-dimensional convolutional encoder, the history buffer by a one-dimensional temporal convolutional encoder, and the scalar vector is concatenated directly. The resulting feature vector is passed to a shared multilayer perceptron (MLP) with separate actor and critic heads.

The policy is trained with proximal policy optimization (PPO; \citealt{2017arXiv170706347S}) using the Stable-Baselines3 implementation \citep{JMLR:v22:20-1364}, with a custom multi-input feature extractor. The production run used 100 million environment steps, 32 parallel environments, rollout length 2048 per environment, mini-batch size 1024, 10 PPO epochs per update, discount factor \(\gamma=0.99\), generalized advantage estimation (GAE) parameter \(\lambda_{\rm GAE}=0.9\), and maximum gradient norm 0.5. The learning rate was warmed up to \(3\times10^{-4}\) during the first 5\% of training and then decayed to \(10^{-5}\). The entropy coefficient was annealed from 0.02 to 0.003.

\subsection{Reward function}
\label{app:rl_reward}

The training reward is the sum of a physically motivated capture term, a potential-based shaping term, and a weak action-smoothness penalty:
\begin{equation}
r_t =
r_t^{\rm cap}
+
r_t^{\rm shape}
+
r_t^{\rm smooth}.
\end{equation}
The capture reward is defined in Equation~\ref{eq:capture_reward}, with $\Delta t=1~{\rm s}$. The time weighting and auxiliary training terms are specified below.

The time weight is chosen to emphasize the most relevant prompt-emission window: it increases linearly from 0.2 to 1.5 during the pre-merger interval \(-60\leq t<0\) s and is set to 3.0 after merger. Both controllers are scored with the same weighted objective in every paired comparison, so the weighting defines the common figure of merit rather than favoring either policy; it concentrates the measured differences in the post-merger window, which is where most of the RL advantage in the reported examples accumulates.

The shaping term is potential-based,
\begin{equation}
r_t^{\rm shape}
=
\gamma_{\rm RL}\Phi(s_{t+1})-\Phi(s_t),
\end{equation}
with
\begin{equation}
\Phi(s_t)
=
\beta
\exp\left(
-\frac{\theta_{\rm peak}(t)}{r_{\rm FoV}}
\right).
\end{equation}
Here \(\gamma_{\rm RL}=0.99\) is the PPO discount factor, \(s_t\) is the environment state at time \(t\), \(\Phi\) is the shaping potential, \(\beta\) is the shaping coefficient, \(\theta_{\rm peak}(t)\) is the angular separation between the telescope pointing and the highest-probability visible pixel in the active sky map, and \(r_{\rm FoV}\) is the effective FoV radius. The coefficient \(\beta\) is annealed from 1 to 0 during training so that the final policy is selected primarily by the capture reward.

The smoothness term is
\begin{equation}
r_t^{\rm smooth}
=
-\lambda_s \left\|\mathbf{a}_t-\mathbf{a}_{t-1}\right\|,
\end{equation}
where \(\mathbf{a}_t\) is the two-dimensional slew action at step \(t\), \(\mathbf{a}_{t-1}\) is the previous action, and \(\lambda_s=0.01\) is the smoothness coefficient. This term discourages rapidly oscillating pointings and is annealed to zero over the first half of training.

For evaluation, we report cumulative unshaped capture reward unless otherwise stated. This ensures that policy comparisons are based on the scientific objective, not on auxiliary training terms.

\subsection{Baselines and evaluation}
\label{app:rl_eval}

The principal comparison is a visibility-aware peak-tracking baseline. At each step, this baseline masks the active HEALPix sky map using the same elevation and declination constraints as the RL environment, identifies the highest-probability visible pixel, and issues the clipped slew command toward that location. This is an intentionally strong baseline for the present problem: for compact or smoothly evolving localizations, following the dominant visible posterior maximum is already close to optimal. However, the peak-tracking baseline is a single-step rule: it always follows the current posterior maximum and does not account for pointing history, future reward weighting, or the finite observing window. 

We evaluate the RL policy and peak-tracking baseline using paired rollouts. For each pair, the event, initial pointing, local sidereal time, visibility geometry, and map realization are held fixed, so that differences in cumulative capture reward reflect the policy rather than the observing geometry. Policy performance is compared using the cumulative unshaped capture reward of the RL policy and the corresponding peak-tracking baseline reward for the same rollout. We denote the cumulative evaluation reward for rollout \(i\) by
\[
J_i = \sum_t r_{t,i}^{\rm cap},
\]
where the sum is over timesteps in that rollout. Across the combined test suite, containing more than $10^4$ paired rollouts, the RL policy closely tracks the visibility-aware peak-tracking baseline. The aggregate return ratio, $\sum_i J_i^{\rm RL}/\sum_i J_i^{\rm base}$, is approximately $0.96$. Using the adopted practical tie threshold, approximately $83\%$ of paired rollouts have effectively equal scores. The remaining cases contain both baseline-favorable and RL-favorable subsets, motivating a classifier that identifies the localizations most likely to benefit from the learned policy.

These results indicate that the learned policy has recovered the main behavior of the visibility-aware peak-tracking baseline in nominal conditions, rather than learning a qualitatively different observing strategy. This result is operationally useful. The peak-tracking controller is simple, interpretable, and easy to validate on real telescope hardware, whereas the RL policy is less transparent because its decisions are produced by a learned actor--critic network. Therefore, a peak-tracking controller is a natural first deployment strategy for prompt DSA follow-up. The learned policy can be used directly for all alerts or selected for individual alerts using the classifier described below.

\subsection{Classifier-gated dispatch of the learned policy}
\label{sec:classifier_dispatch}

Although the two controllers perform similarly in aggregate, there is a small subset of test cases in which the RL policy is more useful. Comparisons selected by the pre-merger classifier described below show a positive aggregate RL advantage at conservative classifier thresholds, although not every selected comparison is an individual RL win. The classifier decision threshold controls this trade-off: a more conservative threshold selects fewer test cases among which the RL advantage is more pronounced. Inspection of this subset suggests a recurring morphology in which the BAYESTAR posterior is broad and asymmetric, with the argmax pixel offset from the probability-mass centroid. Broadness alone is not sufficient to identify an RL-favourable case. For some of these test cases, a field of view centred on the argmax captures only a small fraction of the total probability, so pointing closer to the centroid can be preferable. The learned policy can approximate this behaviour because its training reward is the FoV-integrated probability mass, whereas the peak-tracking controller follows the argmax pixel by construction.

To identify these cases before deployment, we train a lightweight random-forest classifier on five summary features of the earliest available BAYESTAR sky map and current telescope state: the probability captured by a single field of view centred on the argmax pixel, the peak pixel value, the area of the \(90\%\) credible region, the angular offset between the argmax pixel and the posterior-mass centroid, and the angular separation between the telescope's current pointing and the argmax.

\begin{figure}[t]
\centering
\includegraphics[width=\columnwidth]{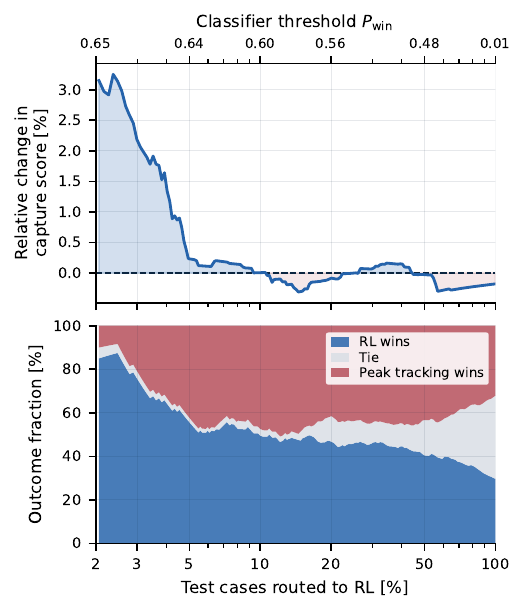}
\caption{Classifier-gated performance as the routing threshold is varied over an exploratory held-out set of approximately \(10^3\) paired test cases. Each test case consists of one simulated event evaluated at a particular observing time and initial telescope pointing. For each classifier threshold, the lower horizontal axis gives the fraction of test cases with \(P_{\rm win}\) above that threshold and therefore routed to RL; the upper axis gives the corresponding empirical \(P_{\rm win}\) threshold. The upper panel shows the aggregate post-merger capture-score change relative to peak tracking within the routed subset, and the lower panel shows the fractions of RL wins, numerical ties, and peak-tracking wins. At \(P_{\rm win}\geq0.65\), approximately \(2\%\) of test cases are routed to RL, yielding a \(3.15\%\) aggregate improvement. }
\label{fig:dispatch_threshold}
\end{figure}

All five features can be computed from the earliest localization and the initial telescope pointing. Given this map and the current telescope pointing, the classifier estimates whether the learned policy or the peak-tracking baseline is more likely to give a higher post-merger capture score. The quantitative trade-off between routing frequency and policy performance is presented in \autoref{fig:dispatch_threshold}.

The RL advantage begins to increase rapidly when routing is restricted to fewer than approximately \(5\%\) of the test cases and becomes substantially stronger for the most conservative selection. At \(P_{\rm win}\geq0.65\), approximately \(2\%\) of held-out test cases are routed to RL.
A possible extension to initial peak-tracking deployment is to use the classifier to select cases for the learned policy before merger. Its operational benefit must be weighed against the additional validation and maintenance of the two controllers and selector. We note that these fractions are measured on noiseless localization sequences (\autoref{sec:simulated_events}). With realistic measurement noise, early-warning posteriors become less regular and their instantaneous maxima less stable. As a consequence, the two controllers disagree more often and the fraction of alerts for which the dispatch decision matters is likely larger than estimated here. A first adaptation of the policy and the dispatch model to Gaussian-noise localizations is presented in \autoref{app:gaussian_noise}.

\subsection{Synthetic map-perturbation stress tests}
\label{app:rl_robustness}

As a diagnostic stress test, we also applied simple synthetic perturbations to all localization stages of a given event, holding the event, observing geometry, and perturbation realization fixed between paired RL and peak-tracking rollouts. These tests include small rigid map rotations, smoothing, log-probability noise, and isolated probability spikes. The two controllers respond similarly to smooth perturbations. Isolated spikes can occasionally pull the peak-tracking controller toward a false local maximum, whereas the RL policy sometimes remains closer to the broader posterior structure or its previous pointing direction. These synthetic tests are intended only to illustrate differences in controller behavior; they are not a substitute for evaluation on noise-realistic localization sequences.

\subsection{Gaussian-noise localization sequences}
\label{app:gaussian_noise}

Relative to the zero-noise sequences, the Gaussian-noise posteriors are broader and less regular, and their instantaneous maxima wander between stages. The same sequences are used for the noisy acquisition fractions quoted in \autoref{sec:reachability}.

Under measurement noise, the dispatch stage requires one modification: a regression model that predicts the expected RL-minus-baseline reward difference from the same five early-map features replaces the binary classifier, and produces a small positive routing gain on each of the confirmation noise realizations, which were excluded from its training. The dispatch decision therefore needs a slightly more expressive model of the same initial-map quantities, rather than additional observables. Real low-latency data contain non-Gaussian, nonstationary features such as glitches; evaluations based on injections into real detector data \citep[e.g.,][]{2024PNAS..12116474C} are a natural next step.

%% For this sample we use BibTeX plus aasjournalv7.bst to generate the

\bibliography{dsa_gw_prompt_radio}{}
\bibliographystyle{aasjournalv7}



\end{document}